# Fisheries Problems and Bureaucracy in Aquaculture Anti-Commons View

**JOSÉ ANTÓNIO FILIPE**
jose.filipe@iscte-iul.pt

**MANUEL ALBERTO M. FERREIRA**
manuel.ferreira@iscte-iul.pt

**MANUEL COELHO**
coelho@iseg.ulisboa.pt

**MARIA ISABEL C. PEDRO**
CEGIST, IST/UTL, *Lisboa, Portugal*
ipedro@ist.ulisboa.pt

**ABSTRACT**

The problems raised by anti-commons and bureaucracy have been linked since the study of Buchanan & Yoon (2000). Bureaucracy involves a multitude of agents that have deciding power. At the view of conflicting interests, the decision makers inertia or the inertia of the system itself, excessive administrative procedures or too many administrative circuits push for too late decisions, or for non-rational decisions in terms of value creation for economic agents. Property Rights Theory explains new concerns. Considering that an "anti-commons" problem arises when there are multiple rights to exclude, the problem of decision process in aquaculture projects makes sense at this level. However, little attention has been given to the setting where more than one person is assigned exclusion rights, which may be exercised. "Anti-commons" problem is analyzed in situations in which resources are inefficiently under-utilized rather than over-utilized as in the familiar commons setting. In this study, fisheries problems are studied and some ways to deal with the problem are presented.

**1.INTRODUCTION**

"Anti-Commons" theory has appeared representing the idea of an excessive partition of property rights. This theory appeared in the 80's of last century, introduced by Michelman (1982). In the last years of the 20$^{th}$ Century several ideas about this new problem around property rights have emerged in which too many rights of exclusion and a reduced level of utilization of the resource are present. Michelman (1982), when presented the notion of "anti-commons", defined it as "a type of property in which everyone always has rights respecting the objects in the regime, and no one, consequently, is ever privileged to use any of them except as particularly authorized by others". This definition of "anti-commons" makes evidence of the lack of efficiency in several situations in which each one of several owners with property rights over a given resource has no effective rights to use the resource (and consequently, each one has the right to exclude other agents from the utilization of the resource).

## 2.PROPERTY RIGHTS DISCUSSION. THE UNDERUSE OF RESOURCES UNDER A SITUATION OF "ANTICOMMONS"

The discussion involving the definition of property rights is old. The types of property rights demand that the limits of these concepts are hardly analyzed. The commons problems have been discussed since the middle of last century, involving the idea that commons reflect usually the overexploitation of resources. The lack of property rights implies that no one may exclude others from access to a given resource. The existence of many agents using a given resource, in these conditions, causes an inefficient level for the resource use and causes a special motivation for agent's over-use the resource. The real level of use for the resource will take place at a higher level compared with the optimal level for society.

On the opposite side, when several owners of a resource have, each one, the right to exclude others from the use of a scarce resource and no one has the full privilege to use it, this resource may have a very limited and unsatisfactory use. This is the problem of the "tragedy of the anti-commons": the resource may be prone to under-use.

After an "anti-commons" emerges, its passage to an efficient process in the private property sphere is long and extremely slow, due to the properties inherent to "anti-commons" and to the difficulties existing for overcoming the "tragedy of anti-commons".

Because of all this, it is necessary to make an important reflection about the definition of property rights to overcome several important aspects when resources are exploited. Indeed, we can see that not just the commons lead to the tragedy. When there are too many property rights and too many rights of exclusion, tragedy seems to be the last result, as well. Too many owners have the right to exclude others but, in fact, none of them have the privilege to use it suitably. Insufficient use is the corollary for this situation.

## 3.AN "ANTI-COMMONS" VIEW

Along this section, it will be taken as a fundamental basis, the work of Schulz, Parisi and Depoorter (2003). They present a general model which permits to distinguish between the simultaneous cases and the sequential cases in "anti-commons" tragedies. As the authors say, the reality may present situations that combine characteristics of the two categories. Anyway, it is important to consider the two situations separately.

For the first of these two cases, they consider that exclusion rights are exercised at the same time, independently. This involves several agents linked in a coincident relationship, such as multiple co-owners with cross-veto powers on the other members' use of the common resource.

In the sequential case, exclusion rights are exercised in consecutive stages, at different levels of the value chain. The several owners of the exclusion rights exercise their own rights in a succession way. Each agent may be involved in a hierarchy and each one may exercise its own exclusion right or veto power over a given proposition (see some examples about simultaneous and sequential anti-commons tragedies in Schulz, Parisi and Depoorter, 2003).

In that work, a dual model of property, where commons and anticommons problems are the consequence of a lack of conformity between use and exclusion rights, is extended to consider the different equilibria obtained under vertical and horizontal cases of property fragmentation.

Horizontal anticommons cases are the ones where situations of exclusion rights exercised simultaneously and independently are present. This involves situations in which are two agents linked in a horizontal relationship. Both agents contribute on the same level of a value chain.

Situations of a vertical anticommons can be expressed as the situations in which the exclusion rights are exercised sequentially by the various right holders. This involves multiple parties in a hierarchy, each of whom can exercise an exclusion or veto power over a given proposition. Examples involving bureaucracy (for example, situations where multiple permits need to be acquired to exercise a given activity) or a production process where a given producer purchases one essential input from a monopolistic seller (see Parisi, Schulz and Depoorter, 2005).

It is shown in Parisi, Schulz and Depoorter (2005) that the symmetrical features of commons and anticommons cases result from a misalignment of the private and social incentives of multiple owners in the use of a common resource. The misalignment is due to externalities not captured in the calculus of interests of the users (commons situations) and excluders (anticommons situations).

The problem of anti-commons is based on a positive externality, when considering it in terms of efficiency[1].

In anti-commons case $x_i$, the activities of agents, shows the extent to which agent $i$ grants agent $j$ permission to use the common property. An activity $x_1$ of agent 1 exerts a positive impact on the productivity of agent 2's activity $x_2$.

If $V_i(x_i, x_j)$ is the value of the common resource of agent $i$, agent $i$ grants agent $j$ the right to use the common resource. Agent $j$ owns a complementary right to exclude agent $i$ from the use of the common resource. These grants are respectively $x_1$ and $x_2$. So, $V_i(x_i, x_j)$ denotes the profit agent $i$ takes from this joint project. The positive externality that agent $j$ exerts on agent $i$ is given by:

---

[3] Commons problem is based on a negative externality.

$$\frac{\partial V_i}{\partial x_j}(x_i, x_j) > 0$$

Assuming that the exclusion rights are exercised simultaneously and independently by the various rights' holders, multiple owners exercise their veto power on equal terms. If both agents are in a perfect symmetric situation,

$$V_i(x_i, x_j) = V_j(x_j, x_i),$$

agent $i$ will be choosing the value of $x_i$ which maximizes $V_i(x_i, x_j)$; and the resulting Nash equilibrium considering agents 1 and 2 is given by

$$\frac{\partial V_1}{\partial x_1}(x_1, x_2) = 0 \text{ and } \frac{\partial V_2}{\partial x_2}(x_2, x_1) = 0.$$

These two conditions are the best response functions of the two agents. Assuming also that $V_i$ is concave in $x_i$, the equilibrium exists for mild assumptions on activities $x_i$. Given the symmetry assumption, a symmetric equilibrium is expected:

$$x_1 = x^c = x_2.$$

There are uncoordinated choices for the 2 agents which can now be compared to the efficient choices of $x_i$, which maximize $V_1 + V_2$ and are characterized by the following first order conditions:

$$\frac{\partial V_1}{\partial x_1}(x_1, x_2) + \frac{\partial V_2}{\partial x_1}(x_2, x_1) = 0 \quad \text{and} \quad \frac{\partial V_2}{\partial x_2}(x_2, x_1) + \frac{\partial V_1}{\partial x_2}(x_1, x_2) = 0.$$

Given the symmetric assumption, a symmetric optimum is expected. Assuming $V_1 + V_2$ concave and that there is a symmetric solution, the efficient choices are equal:

$$x_1 = x^s = x_2.$$

As can be easily concluded, $x^s > x^c$. This is, the uncoordinated choices lead to the underutilization of the common resource.

If horizontal and vertical anticommons are introduced, both being obviously the consequence of the existence of non-conformity between use and exclusion rights, the problem of underutilization of resources is exacerbated if the right is fragmented into more than two exclusion rights, with more than two agents deciding independently on their activity or prize (see Schulz, 2000 and Parisi, Schulz and Depoorter (2005).

## 4. TRAGEDY OF FISHING COMMONS. LOOKING FOR SOLUTIONS

The classical theory of the "tragedy of the commons" explains the reasons why sea fisheries are prone to over-exploitation (see Hardin, 1968). Gordon (1954) had already examined the problems of common resources in the 50s.

The emergence of individual transferable quotas (ITQs) would potentially allow overcoming, even partially, the serious problem of over-fishing for several species in several parts of the world. Anyway, the enormous enthusiasm around ITQs has given place to the appearance of the problem of the "anti-commons".

Some nations have tried to avoid the "tragedy of the commons" in sea fisheries by regulating the activity, decade over decade. Reducing fishing seasons, restricting open areas to fisheries, limiting the use of gears or reducing the power and tons for ships were just some measures to avoid tragedies to species. The truth is that these practices many times did not reduce over-fishing for the species that governments intended to protect. Fishing race and massive discards, frequently, continued.

TACs (total allowable catches) are fixed each year by fishing management authorities. Each fisher has a part of the TAC to fish, representing his own individual quota. Theoretically, with this procedure, each fisher may use his quota when he likes and fishing race may come to an end (see Leal, 2002a). Meanwhile, quotas may be transferable, and this procedure may lead to situations in which the quotas owners can adjust the dimension of his fishing operations buying or selling quotas or even just leaving the fishery and move the quota from the market.

Many nations have been using this kind of measure (programs of individual transferable quotas) to manage fishing resources in their waters. These programs contributed to improving fishers' rents, to improve the quality of products, to reduce the excess of quotas and to eliminate eventual catches that exceed TACs (see Alesi, 1998; see Repetto, 2001 and Wilen and Homans, 2000, as well).

## 5. THE EMERGENCE OF ANTI-COMMONS. ALASKA'S HALIBUT EXAMPLE

Alaska's Halibut allows to study the effect of the existence of ITQs and, additionally, to study the consequences in terms of the "anti-commons" (see Leal, 2002b). In fact, this specie got overexploited, and authorities implemented several measures to reduce catches. First, fishing seasons were reduced. At the beginning of the 90's, fisheries were opened just for two or three short periods of about 24 hours per year. Consequently, fishing race became the solution for fishers, who tried to get the maximum fish as possible, throughout the available time for fishing. In fact, results were different than the expected ones for the worse. However, after the implementation of individual quotas in 1995, fishing seasons became larger and fishers could exploit this resource for around 8 months, per year. Sales increased and prices got higher (see GAO, 2002). Meanwhile, catches got smaller than TACs and fleets excesses were reduced.

Nevertheless, individual quotas excess may lead to sub-exploitation of the resource and Alaska's Halibut is, in fact, a case that must be studied to see the consequences of too many existing fishing quotas. Authorities have implemented rules to protect small fishers. They did not authorize fishers to sell their quotas if they were very small. Consequently, these quotas went unexploited, because they were not profitable for their owners. Halibut got underexploited. Authorities had to change the rules for this fishery. It can be seen now that ITQ promote important solutions for the "tragedy of the commons", but they may create conditions for a new situation of "anti-commons".

Some other examples may be presented. For example, Leal (2002b) shows how Alaska's crab may be prone to under-use if fishers are forced to sell their catches to a little number of companies, as it was the case. Low prices lead to situations in which fishers under-use their quotas because they got unprofitable. As a consequence, crab got under-exploited.

### 6.THE AQUACULTURE CASE IN PORTUGAL

The example of aquaculture in Portugal can be presented. There are too many entities to analyze projects of aquaculture. Rules and procedures are so many that projects are approved with long delays[2]. Therefore, resources get underused.

The aquaculture sector in Portugal is studied and allows to evaluate the possibility of using the hypothesis suggested by Buchanan & Yoon (2000) that bureaucracy can be studied with the help of the anti-commons conceptualization.

In this context, some questions are posed about live resources exploitation, particularly in fisheries and aquaculture projects and raised the legal problems in the Portuguese case. An economic analysis allows to show how this problem of anti-commons can originate an important loss of value. It is seen as anti-commons tragedies appear in such situations in the aquaculture problem.

The suggestion of Buchanan and Yoon (2000) that the anti-commons construction offers an analytical means of isolating a central feature of "sometimes disparate institutional structures" shows the problems arisen from bureaucracy in this context. The persistence of bureaucratic circuits of approval and implementation of projects can make the entrepreneurship activities and it diminishes the potential of regional and coastal development.

The responsible Department for Aquaculture is specifically DGPA (Direcção Geral das Pescas e Aquicultura), which is responsible for supervising and controlling the activity of aquaculture sector[3]. There are an enormous set of initial steps for a project's approval (Decreto Regulamentar nº 14/2000) and there are many entities deciding (see

---

[2] Other countries have similar problems in this kind of projects.

[3] The aquaculture problem is fitted under the control and supervision of Ministério da Agricultura, do Desenvolvimento Rural e das Pescas (see Decreto Regulamentar nº 14/2000 – September, 21st, 2000), that is the Ministry for Agriculture, Fishing and Aquaculture Sector. This Decree specifies the requisites and conditions needed to install and exploit a plant on this area. The Decreto Regulamentar nº 9/2008 (March 18th, 2008) defines a set of rules specifically for installations offshore.

Filipe *et al*, 2011a). This leads to projects rejection or a very delayed approval. Besides it shows how bureaucracy is involved in worsening the conditions of exploitation of the resources. The "disparate institutional structures" get evident and a problem of anti-commons is the obvious result.

# 7.CONCLUDING REMARKS

The anti-commons framework has permitted very interesting studies on several areas of resources. It is shown in this study how anti-commons are associated with negative externalities and to a underuse of resources.

Some examples in fisheries are presented. The aquaculture case for projects in Portugal is also studied. Aquaculture contributes for fish production and being projects in this area ecologically sustainable, as they often intend, they may contribute for solving fisheries' dilemmas about sea fishing resources exploitation without creating other environment problems.

In Portugal an excessive number of regulators (some of them with veto power) analyze the projects. They spent too much time to overpass all the steps and when the process is ready for implementation it may be too late (and sometimes, the project is refused).

Too many resources are spent on projects and they simply become unviable. A project may create value for the investor and for the community, but all the time wasted in bureaucratic analysis makes the project unviable.